\documentclass[useAMS,usenatbib,letterpaper]{mn2e}
\usepackage{graphicx,rotating,subfigure,amsmath,wasysym,amssymb}
\usepackage[retainorgcmds]{IEEEtrantools}

\title[Continuum removal in H$\alpha$ measurements]
      {Continuum removal in H$\alpha$ extragalactic measurements}
\author[O. Spector, I. Finkelman and N. Brosch]
{O. Spector$^{1}$\thanks{E-mail: odedspec@wise.tau.ac.il}, I. Finkelman and N. Brosch \\
$^{}$Wise Observatory and the Beverly and Raymond Sackler School of Physics and Astronomy, \\
       Tel Aviv University, Tel Aviv 69978, Israel\\
}

\begin{document}

\date{Accepted 2011 September 21. Received 2011 September 21; in original form 2011 August 17}

\pagerange{\pageref{firstpage}--\pageref{lastpage}} \pubyear{2011}

\maketitle

\label{firstpage}

\begin{abstract}
We point out an important source of error in measurements of extragalactic H$\alpha$ emission and suggest ways to reduce it. 

The H$\alpha$ line, used for estimating star formation rates, is commonly measured by imaging in a narrow band and a wide band, both which include the line. The image analysis relies on the accurate removal of the underlying continuum. 
We discuss in detail the derivation of the emission line's equivalent width and flux for extragalactic extended sources, and the required photometric calibrations. We describe commonly used continuum-subtraction procedures, and discuss the uncertainties that they introduce. 

Specifically, we analyse errors introduced by colour effects.
We show that the errors in the measured H$\alpha$ equivalent width induced by colour effects can lead to underestimates as large as 40\% and overestimates as large as 10\%, depending on the underlying galaxy's stellar population and the continuum-subtraction procedure used.
We also show that these errors may lead to biases in results of surveys, and to the underestimation of the cosmic star formation rate at low redshifts (the low z points in the Madau plot). 
We suggest a method to significantly reduce these errors using a single colour measurement.
\end{abstract}


\section{Introduction}
Physical properties of galaxies can often be derived from spectroscopic observations, which allow the detection and characterisation of various emission lines.
However, the complexities related with observation and reduction procedures, make spectroscopic surveys considerably more difficult to pursue than imaging surveys using narrow-band filters. 
While such imaging surveys detect only lines whose redshifted wavelength fall inside the filter's band, they have the advantage of detecting spatially extended line emission down to low surface brightness levels. 
This imaging technique is, therefore, useful for evaluating the full spatial extent and distribution of the line-emission in nearby galaxies (Kennicutt et al.\ 2008), as well as to search and identify high-redshift line-emission galaxies in studies where many objects are targeted (e.g., Thompson, Mannucci \& Beckwith 1996; Nilsson et al.\ 2007).

The H$\alpha$ line provides a large part of our knowledge of star formation (SF) activity in galaxies and is the focus of this paper.
Observing a galaxy with a filter centred at rest-frame H$\alpha$ can, with reasonable assumptions, provide information on the amount and spatial distribution of the emitted ionising radiation using relatively modest integration times on small telescopes, thus making large galaxy surveys practical.

Since the flux measured by narrow-band imaging includes both the spectral line and the continuum emission, it is necessary to measure the continuum contribution. In many cases this is done by imaging the target in a band significantly wider than the narrow band (e.g., Waller 1990; James et al.\ 2004; Kennicutt et al.\ 2008; 	 Brosch, Spector \& Zitrin 2011). A scaling factor between the continuum counts of the wide and narrow band images (Wide to Narrow Continuum Ratio - WNCR) is then determined. From its value the line's counts can be extracted.

This paper discusses the derivation of emission line properties from wide and narrow band imaging, focusing on sources of measurement errors.
In Section~\ref{sec:Derivations_EW} we derive the relations required to calculate the equivalent width (EW) and the flux. 
Section~\ref{sec:photometry} discusses methods for obtaining a calibrated measure of the flux. 
In Section~\ref{sec:Err_Propagation} we derive the effect of the WNCR estimation error on the measurement errors of the line's equivalent width and flux.
In Section~\ref{sec:WNCR_MeasMethod} we review some commonly used methods to measure WNCR. Then, in Section~\ref{sec:ColourEffects} we analyse the measurement errors that originate from colour effects when using these methods, and quantify these errors for the case of H$\alpha$ measurements in the local Universe.

\section{Derivation of Emission-Line Properties}
\label{sec:Derivations_EW}
Line emission from a galaxy or a region within a galaxy (target) can be measured by observing it through a narrow-band filter (N) and a wide-band filter (W), where the N band contains the emission-line, and the W band may or may not contain it. In this section we derive the method to calculate the emission-line's equivalent width (EW) and flux ($\mbox{F}_{line}$) from such observations, and discuss the different assumptions made in this derivation and their implications. The derivation and discussion are general, and can be applied to H$\alpha$ measurements as well as to other emission lines measured in a similar way.

The response of the electro-optical system can be expressed as the integral of the observed flux density over wavelength, weighted by the transmittance functions of the filter and of the atmosphere and by the response function of the rest of the electro-optical system:
\begin{equation}
\begin{IEEEeqnarraybox*}{rCl} 
\mbox{cps}_{N} & = & \int_0^\infty \mbox{F}_\lambda \: \mbox{T}_{atm,N}(\lambda) \: \mbox{T}_{N}(\lambda) \: \mbox{R}_{\lambda} \: d\lambda \\
\mbox{cps}_{W} & = & \int_0^\infty \mbox{F}_\lambda \: \mbox{T}_{atm,W}(\lambda) \: \mbox{T}_{W}(\lambda) \: \mbox{R}_{\lambda} \: d\lambda 
\end{IEEEeqnarraybox*} 
\end{equation}
where:
\begin{description}
  \item[$\mbox{cps}_{N}$, $\mbox{cps}_{W}$] are the measured count rates of the narrow-band (N) and wide-band (W) filters (respectively) in instrumental units - typically ADU s$^{-1}$ (Analog to Digital Units per second);
  \item[$\mbox{F}_\lambda$] is the spectral flux density of light from the observed target reaching the telescope in absence of an atmosphere (typically in erg~s$^{-1}$~cm$^{-2}$~$\mbox{\AA}^{-1}$);
  \item[$\mbox{T}_{N}(\lambda)$, $\mbox{T}_{W}(\lambda)$] are the unitless transmittance functions of the N and W bands, respectively, ranging from 0 (fully opaque) to 1 (fully transparent);
  \item[$\mbox{T}_{atm,N}(\lambda)$, $\mbox{T}_{atm,W}(\lambda)$] are the atmospheric transmittance as function of wavelength, including effects of weather, elevation and airmass of observation, when the N and W bands (respectively) were imaged (unitless, ranging from 0 to 14);
  \item[$\mbox{R}_{\lambda}$] is the responsivity as function of wavelength of the rest of the electro-optical system (i.e. the telescope and sensors, excluding the transmittance effect of the filters) typically in ADU~erg$^{-1}$~cm$^{2}$;
  \item[$\lambda$] is the wavelength (typically in $\mbox{\AA}$).
  \item[]
\end{description}

This paper considers the case where the measured line is the only significant spectral feature in the N and W bands, i.e. that the target's flux can be divided into line and continuum (cont) emission. For the case of H$\alpha$, this means that some adjustments should be made, if the [NII] lines flanking it carry significant flux.

The count rate from the observed target can be expressed as the sum of the line and continuum contributions:
\begin{equation} 
\begin{IEEEeqnarraybox*}{rCl} 
\mbox{cps}_{N} & = & \mbox{cps}_{N,cont}+\mbox{cps}_{N,line} \\
\mbox{cps}_{W} & = & \mbox{cps}_{W,cont}+\mbox{cps}_{W,line}
\end{IEEEeqnarraybox*}
\label{cps_cont_line} 
\end{equation}
\\
\begin{equation} 
\begin{IEEEeqnarraybox*}{lCl}
\mbox{cps}_{N,cont} & = & \int_0^\infty \mbox{F}_{\lambda,cont} \: \mbox{T}_{atm,N}(\lambda) \: \mbox{T}_{N}(\lambda) \: \mbox{R}_{\lambda} \: d\lambda \\ 
\mbox{cps}_{N,line} & = & \int_0^\infty \mbox{F}_{\lambda,line} \: \mbox{T}_{atm,N}(\lambda) \: \mbox{T}_{N}(\lambda) \: \mbox{R}_{\lambda} \: d\lambda \\
\mbox{cps}_{W,cont} & = & \int_0^\infty \mbox{F}_{\lambda,cont} \: \mbox{T}_{atm,W}(\lambda) \: \mbox{T}_{W}(\lambda) \: \mbox{R}_{\lambda} \: d\lambda \\
\mbox{cps}_{W,line} & = & \int_0^\infty \mbox{F}_{\lambda,line} \: \mbox{T}_{atm,W}(\lambda) \: \mbox{T}_{W}(\lambda) \: \mbox{R}_{\lambda} \: d\lambda
\end{IEEEeqnarraybox*}
\label{cps_cont_line_detailed}
\end{equation}
where:
\begin{description}
  \item[$\mbox{cps}_{N,line}$, $\mbox{cps}_{N,cont}$] are respectively the line and continuum contributions to $\mbox{cps}_{N}$; 
  \item[$\mbox{cps}_{W,line}$, $\mbox{cps}_{W,cont}$] are respectively the line and continuum contributions to $\mbox{cps}_{W}$;
  \item[$\mbox{F}_{\lambda,line}$, $\mbox{F}_{\lambda,cont}$] are respectively the line and continuum contributions to $\mbox{F}_\lambda$.
  \item[]
\end{description}

Since the width of the emission-line is typically very narrow in comparison to both the N and W bands, its profile can be approximated as a delta function at the centre of the emission line:
\begin{equation}
\mbox{F}_{\lambda,line} \approx \mbox{F}_{line} \: \delta(\lambda - \lambda_{line})
\end{equation}
where:
\begin{description}
  \item[$\lambda_{line}$] is the central wavelength of the emission line (typically in $\mbox{\AA}$);
  \item[$\mbox{F}_{line} = \int_0^\infty \mbox{F}_{\lambda,line} \: d\lambda$] - the flux (irradiance) of the emission line (typically in erg~s$^{-1}$~cm$^{2}$).
  \item[]
\end{description}

Using this approximation, the line contribution is simply: 
\begin{equation}
\begin{IEEEeqnarraybox*}{rCl} 
\mbox{cps}_{N,line} &\cong&  \mbox{F}_{line} \: \mbox{T}_{atm,N}(\lambda_{line}) \: \mbox{T}_{N}(\lambda_{line}) \: \mbox{R}_{\lambda}(\lambda_{line}) \\
\mbox{cps}_{W,line}  &\cong&  \mbox{F}_{line} \: \mbox{T}_{atm,W}(\lambda_{line}) \: \mbox{T}_{W}(\lambda_{line}) \: \mbox{R}_{\lambda}(\lambda_{line})
\end{IEEEeqnarraybox*} 
\label{cps_line_approx}
\end{equation} 

The goal of the measurement is to find the emission-line's flux, $\mbox{F}_{line}$, and equivalent width, EW.
The line's equivalent width is defined as:
\begin{equation}
\mbox{EW} \equiv \frac { \mbox{F}_{line}} { \mbox{F}_{\lambda,cont}(\lambda_{line}) }
\label{EW_definition}
\end{equation}

while, the flux can be extracted from \eqref{cps_line_approx}:
\begin{equation}
\mbox{F}_{line}  \cong  \frac { \mbox{cps}_{N,line} } 
                              { \mbox{T}_{atm,N}(\lambda_{line}) \: \mbox{T}_{N}(\lambda_{line}) \:
                                \mbox{R}_{\lambda}(\lambda_{line}) }
\label{F_line}
\end{equation}

The term $\mbox{T}_{atm,N}(\lambda_{line}) \cdot \mbox{R}_{\lambda}(\lambda_{line})$ is obtained by photometric calibrations and is discussed in section~\ref{sec:photometry}. The N-band's transmittance, $\mbox{T}_{N}(\lambda_{line})$, measured directly or provided by the filter manufacturer, should be corrected to compensate for effects of the telescope's converging beam, since the transmittance curves of interference filters (the type of filters usually used) vary with angle of incidence. 

We now derive the remaining terms: $\mbox{cps}_{N,line}$ and $\mbox{F}_{\lambda,cont}(\lambda_{line})$.
Using the two parts of \eqref{cps_line_approx}, we can write $\mbox{cps}_{W,line}$ in terms of $\mbox{cps}_{N,line}$ (based on the $\delta$ function line approximation):
\begin{equation}
\mbox{cps}_{W,line}  \cong   \mbox{cps}_{N,line} \: 
                             \frac{ \mbox{T}_{atm,W}(\lambda_{line}) \: \mbox{T}_{W}(\lambda_{line}) }
                                  { \mbox{T}_{atm,N}(\lambda_{line}) \: \mbox{T}_{N}(\lambda_{line}) }
\end{equation}

Using this and \eqref{cps_cont_line} we can express $\mbox{cps}_{W,cont}$ as:  

\begin{equation}
\begin{IEEEeqnarraybox*}{lCl}
\mbox{cps}_{W,cont} & \cong & \mbox{cps}_{W} - \mbox{cps}_{N,line} \: 
                              \frac{ \mbox{T}_{atm,W}(\lambda_{line}) \: \mbox{T}_{W}(\lambda_{line}) }
                                   { \mbox{T}_{atm,N}(\lambda_{line}) \: \mbox{T}_{N}(\lambda_{line}) } \\
\end{IEEEeqnarraybox*}
\label{cps_cont_W}
\end{equation}

For the derivation of $\mbox{cps}_{N,line}$ we define the Wide to Narrow Continuum Ratio (WNCR) to be the ratio between the count rate contributed by the continuum in the W band and the count rate contributed by the continuum in the N band:
\begin{equation}
\mbox{WNCR} \\ \equiv \frac { \mbox{cps}_{W,cont} } { \mbox{cps}_{N,cont} } 
\label{WNCR_def}
\end{equation}

In sections \ref{sec:Err_Propagation}, \ref{sec:WNCR_MeasMethod} and \ref{sec:ColourEffects} we describe the methods most commonly used to measure WNCR, and analyse their effect on the accuracy of EW and $\mbox{F}_{line}$ measurements.

From \eqref{cps_cont_line}, \eqref{cps_cont_W} and \eqref{WNCR_def} we obtain an expression for $\mbox{cps}_{N,line}$:
\begin{equation}
\begin{IEEEeqnarraybox*}{lCl}
\mbox{cps}_{N,line} & = & \mbox{cps}_{N} - \mbox{cps}_{N,cont} 
                      =   \mbox{cps}_{N} - \frac{ \mbox{cps}_{W,cont} }{ \mbox{WNCR} } \\
                 & \cong & \mbox{cps}_{N} - \frac
                                { \mbox{cps}_{W} - \mbox{cps}_{N,line} \: 
                                  \frac{ \mbox{T}_{atm,W}(\lambda_{line}) \: \mbox{T}_{W}(\lambda_{line}) } 
                                       { \mbox{T}_{atm,N}(\lambda_{line}) \: \mbox{T}_{N}(\lambda_{line}) } 
                                }
                                { \mbox{WNCR} }
\end{IEEEeqnarraybox*}
\end{equation}

Extracting $\mbox{cps}_{N,line}$ yields:

\begin{equation}
\begin{IEEEeqnarraybox*}{lCl}
\mbox{cps}_{N,line} 
    & \cong &
        \left(\mbox{cps}_{N} - \frac { \mbox{cps}_{W} }{ \mbox{WNCR} } \right) \cdot  \\
    &&  \left[1 -  \frac{1}{\mbox{WNCR}} \cdot 
                   \frac{ \mbox{T}_{atm,W}(\lambda_{line}) \: \mbox{T}_{W}(\lambda_{line}) } 
                        { \mbox{T}_{atm,N}(\lambda_{line}) \: \mbox{T}_{N}(\lambda_{line}) } 
        \right]^{-1}
\end{IEEEeqnarraybox*}
\label{cps_line_N_solved}
\end{equation}

Note that this expression includes the atmospheric transmittance at $\lambda_{line}$ for both bands: $\mbox{T}_{atm,W}(\lambda_{line})$ and $\mbox{T}_{atm,N}(\lambda_{line})$. These can be estimated from a photometric calibration in the N band, when available. If the W and N exposures were taken at very similar airmasses and atmospheric conditions, these can be approximated as equal  $\left( \frac{\mbox{T}_{atm,W}(\lambda_{line})}{\mbox{T}_{atm,N}(\lambda_{line})} \cong 1 \right)$.

Typically, the N band is chosen so that $\mbox{T}_{N}(\lambda_{line})$ is as high as possible ($\lambda_{line}$ is close to the N band's peak), and is not significantly smaller than $\mbox{T}_{W}(\lambda_{line})$. The W band is typically chosen to be significantly wider than the N band, therefore $ \mbox{WNCR} \gg 1$, and also $\frac{ \mbox{T}_{W}(\lambda_{line}) }{ \mbox{WNCR} \cdot \mbox{T}_{N}(\lambda_{line}) } \ll 1$. This means that the error due to small deviations from the above approximation will typically be minor. It will decrease the closer $\lambda_{line}$ is to the N band's peak and the wider the W band is compared to the N band.

In order to derive $\mbox{F}_{\lambda,cont}(\lambda_{line})$ we approximate $\mbox{F}_{\lambda,cont}$ to be constant within the N band. This is valid if the N filter is very narrow, and has negligible ``out of band leaks'' (a requirement which is not trivial for very narrow bands). If $\mbox{F}_{\lambda,cont}$ is constant within the N band, it can be taken out of the integral for $\mbox{cps}_{N,cont}$ in \eqref{cps_cont_line_detailed} and be replaced by $\mbox{F}_{\lambda,cont}(\lambda_{line})$:

\begin{equation}
\mbox{cps}_{N,cont} 
  \cong  
\mbox{F}_{\lambda,cont}(\lambda_{line}) \: 
   \int_0^\infty \mbox{T}_{atm,N}(\lambda) \: \mbox{T}_{N}(\lambda) \: \mbox{R}_{\lambda} \: d\lambda
\label{cpsN_cont_est}
\end{equation}

Using \eqref{cps_cont_line}, we get:

\begin{equation}
\mbox{F}_{\lambda,cont}(\lambda_{line}) \cong  
   \frac{\mbox{cps}_{N} - \mbox{cps}_{N,line}}
        {\int_0^\infty \mbox{T}_{atm,N}(\lambda) \: \mbox{T}_{N}(\lambda) \: \mbox{R}_{\lambda} \: d\lambda}
\label{F_cont_lambda_solved}
\end{equation}

Replacing $\mbox{F}_{\lambda,cont}(\lambda_{line})$ and $\mbox{F}_{line}$ in \eqref{EW_definition} with the expressions from \eqref{F_cont_lambda_solved}, and using equation \eqref{F_line} yields:

\begin{equation}
\mbox{EW} \cong  \frac { \mbox{cps}_{N,line} } 
                {\mbox{cps}_{N} - \mbox{cps}_{N,line}}
          \cdot
          \frac {\int_0^\infty \mbox{T}_{atm,N}(\lambda) \: \mbox{T}_{N}(\lambda) \: \mbox{R}_{\lambda} \: d\lambda}
                { \mbox{T}_{atm,N}(\lambda_{line}) \: \mbox{T}_{N}(\lambda_{line}) \: \mbox{R}_{\lambda}(\lambda_{line}) } 
\end{equation}

If the N band is sufficiently narrow, $\mbox{R}_{\lambda}$ and $\mbox{T}_{atm,N}$ can be assumed constant within it. Thus:

\begin{equation}
\mbox{EW} \cong  \frac { \mbox{cps}_{N,line} } 
                {\mbox{cps}_{N} - \mbox{cps}_{N,line}}
          \cdot
          \frac {\int_0^\infty \mbox{T}_{N}(\lambda) \: d\lambda}
                { \mbox{T}_{N}(\lambda_{line}) } 
\label{EW_solved}
\end{equation}

The equivalent width, EW, can therefore be obtained from \eqref{cps_line_N_solved} and \eqref{EW_solved} without requiring photometric calibrations. It is equal to the ratio between the count rates of the line ($\mbox{cps}_{N,line}$) and the continuum ($\mbox{cps}_{N} - \mbox{cps}_{N,line}$) multiplied by the ``N-band effective width'' $\left( \frac { \int_0^\infty \mbox{T}_{N}(\lambda) \: d\lambda } { \mbox{T}_{N}(\lambda_{line}) } \right)$. The ``N-band effective width'' is a measure of the N-band width multiplied by the ratio between the N-band peak transmittance and the N-band transmittance at the emission line's wavelength, $\mbox{T}_{N}(\lambda_{line})$. 

The EW measurement error contributed by atmospheric effects originates only from $\mbox{cps}_{N,line}$. In order to understand how changes in the atmospheric transmittance (e.g., clouds) affect the accuracy of the result, we write \eqref{cps_line_N_solved} with WNCR in the last term expressed through fundamental variables, as defined in \eqref{cps_cont_line_detailed} and \eqref{WNCR_def}:

\begin{equation}
\begin{IEEEeqnarraybox*}{lCl}
\mbox{cps}_{N,line} 
     && \cong 
        \left(\mbox{cps}_{N} - \frac { \mbox{cps}_{W} }{ \mbox{WNCR} } \right) \cdot  \\
     &&  \left[1 -  \frac 
          { \int_0^\infty \mbox{F}_{\lambda,cont} \: \frac{\mbox{T}_{atm,N}(\lambda)}{\mbox{T}_{atm,N}(\lambda_{line})} \:
            \frac{\mbox{T}_{N}(\lambda)}{\mbox{T}_{N}(\lambda_{line})} \: \mbox{R}_{\lambda} \: d\lambda 
          }
          { \int_0^\infty \mbox{F}_{\lambda,cont} \: \frac{\mbox{T}_{atm,W}(\lambda)}{\mbox{T}_{atm,W}(\lambda_{line})} \:
            \frac{\mbox{T}_{W}(\lambda)}{\mbox{T}_{W}(\lambda_{line})} \: \mbox{R}_{\lambda} \: d\lambda 
          }
        \right]^{-1}
\end{IEEEeqnarraybox*}
\end{equation}

It can be seen here and in \eqref{EW_solved} that atmospheric conditions affect EW only through the relative transmissions: $\frac{\mbox{T}_{atm,N}(\lambda)}{\mbox{T}_{atm,N}(\lambda_{line})}$ and $\frac{\mbox{T}_{atm,W}(\lambda)}{\mbox{T}_{atm,W}(\lambda_{line})}$. ``Grey'' wavelength-independent effects do not change these terms, and therefore will not affect the result. 

The effect of clouds blocking the field during parts of the exposure will be the same as reducing $\mbox{T}_{atm,N}(\lambda)$ and/or $\mbox{T}_{atm,W}(\lambda)$ by an unknown factor. This effect is virtually independent of wavelength, and therefore does not contribute to the measurement error in EW, whether the W and N exposures were taken in similar airmass and atmospheric conditions or not.

\section{Photometric Calibrations} 
\label{sec:photometry}

Unlike EW, measuring the line's flux, $\mbox{F}_{line}$, requires a photometric calibration. $\mbox{F}_{line}$ can be obtained from \eqref{F_line}, if $\mbox{T}_{atm,N}(\lambda_{line}) \cdot \mbox{R}_{\lambda}(\lambda_{line})$ is known. A useful method to measure \mbox{$\mbox{T}_{atm,N}(\lambda_{line}) \cdot \mbox{R}_{\lambda}(\lambda_{line})$} utilises photometric calibration in the N band. Spectrophotometric standard stars (i.e. stars with known $\mbox{F}_\lambda$) should be measured at several airmass (AM) values. The measured count rate of the standard stars can be expressed as:

\begin{equation}
\mbox{cps}_{N,spec*}\left(AM\right)  =  \int_0^\infty \mbox{F}_{\lambda,spec*} \: \mbox{T}_{atm,cal}(\lambda,AM) \: \mbox{T}_{N}(\lambda) \: \mbox{R}_{\lambda} \: d\lambda \\
\end{equation}
where:
\begin{description}
  \item[$\mbox{cps}_{N,spec*}\left(AM\right)$] is the count rate of the spectrophotometric standard star imaged through the narrow-band (N) filter as function of the airmass (typically in ADU s$^{-1}$);
  \item [$\mbox{F}_{\lambda,spec*}$] is the spectral flux density of the spectrophotometric standard star;
  \item [$\mbox{T}_{atm,cal}(\lambda,AM)$] is the atmospheric transmittance during the photometric calibration as function of wavelength and airmass.
  \item[]
\end{description}

If the N band is sufficiently narrow, $\mbox{R}_{\lambda}$ and $\mbox{T}_{atm,cal}$ can be approximated as constant within the N band. This yields:
\begin{equation}
\mbox{T}_{atm,cal}(\lambda_{line},AM) \cdot \mbox{R}_{\lambda}(\lambda_{line}) =  
    \frac {\mbox{cps}_{N,spec*}\left(AM\right)}
          {\int_0^\infty \mbox{F}_{\lambda,spec*} \: \mbox{T}_{N}(\lambda)  \: d\lambda} \\
\end{equation}

From measurements at different airmasses, a fit can be made to the following logarithmic relation:
\begin{equation}
-2.5 \: \mbox{log} \left[ \mbox{T}_{atm,cal}(\lambda_{line},AM) \cdot \mbox{R}_{\lambda}(\lambda_{line}) \right] 
   \cong 
\mbox{AM} \cdot \mbox{E}_{\lambda_{line}} + \mbox{M}_0
\label{Tatm_fit}
\end{equation}
where:
\begin{description}
  \item[$\mbox{E}_{\lambda_{line}}$] is the extinction coefficient of the atmosphere at $\lambda_{line}$;
  \item[$\mbox{M}_0$] is the zero point of the fit.
  \item[]
\end{description}

If the N band image is taken at similar atmospheric conditions as during the calibration, and with AM for which the fit \eqref{Tatm_fit} still holds, $\mbox{T}_{atm,N}(\lambda_{line}) \cdot \mbox{R}_{\lambda}(\lambda_{line})$ can be simply obtained by:

\begin{equation}
\mbox{T}_{atm,N}(\lambda_{line}) \cdot \mbox{R}_{\lambda}(\lambda_{line}) 
  \cong 
10 ^ {-0.4 \left( \mbox{AM}_N \cdot \mbox{E}_{\lambda_{line}} + \mbox{M}_0 \right) }
\end{equation}
where:
\begin{description}
  \item[$\mbox{AM}_N$] is the airmass through which the N band image was taken.
  \item[]
\end{description}

When photometric calibrations of the N band cannot be obtained, one can calibrate the emission-line's flux based on wide-band calibrated images. The target has to be measured through several calibrated wide bands, to obtain its flux per wavelength, $\mbox{F}_\lambda$, averaged over each band. From these averaged $\mbox{F}_\lambda$ values a model for the target's continuum flux as function of wavelength, $\mbox{F}_{\lambda,cont}$, can be derived. Using \eqref{EW_definition} with the modelled $\mbox{F}_{\lambda,cont}(\lambda_{line})$, and the measured EW, the flux can be simply obtained: 

\begin{equation}
\mbox{F}_{line}  =  \mbox{EW} \cdot \mbox{F}_{\lambda,cont}(\lambda_{line})
\label{F_line_solved}
\end{equation}

When available, wide band magnitudes from the Sloan Digital Sky Survey (SDSS - Abazajian et al. 2009) can be used to model $\mbox{F}_{\lambda,cont}$. In some cases a linear fit of the $g$, $r$ and $i$ magnitudes (whose average wavelengths are 4686$\mbox{\AA}$, 6165$\mbox{\AA}$, and 7481$\mbox{\AA}$ respectively) is a sufficient approximation for the continuum at H$\alpha$. The error of this linear approximation can be estimated using the residuals of the fit.

\begin{figure}

\includegraphics[width=8cm,trim=0mm 0mm 0mm 0, clip]{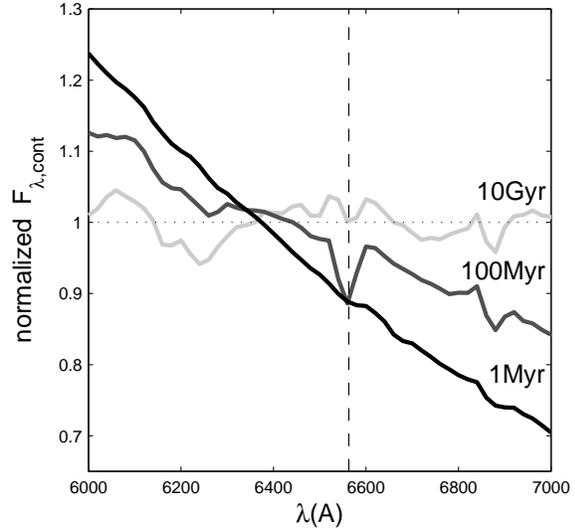}

\caption{$\mbox{F}_{\lambda,cont}$ of $1$, $100$\,Myr and $10$\,Gyr old SSPs at z=0, normalized so that the average $\mbox{F}_{\lambda,cont}$ over the R-band of each stellar population would be 1. Data adopted from the Bruzual-Charlot atlas available in ftp://ftp.stsci.edu/cdbs/grid/bc95/templates/ (also used for the calibration of several {\it Hubble Space Telescope} instruments). The dotted horizontal line shows the model adopted, when calibrating using the R-band alone ($\mbox{F}_{\lambda,cont}$ constant over the R band). The dashed vertical line indicates the H$\alpha$ wavelength ($\lambda_{H\alpha}$).
\label{f:cont}}
\end{figure}

Since typically the emission-line's wavelength is within the W band, one might be tempted to approximate $\mbox{F}_{\lambda,cont}(\lambda_{line})$ to be the average continuum flux measured for W (when a photometric calibration for band W is available). Such an approximation is equivalent to a model in which $\mbox{F}_{\lambda,cont}$ is constant within the W band, and is therefore highly inaccurate as evident from Figure \ref{f:cont}.

Figure \ref{f:cont} shows three examples of normalized $\mbox{F}_{\lambda,cont}$ curves ($1$\,Myr, $100$\,Myr and $10$\,Gyr old stellar populations). The na\"{\i}ve model of constant $\mbox{F}_{\lambda,cont}$ over the R band is also shown (dotted horizontal line). Using a photometric calibration for W alone (R in this example) implies that the dashed line is used instead of the real value. For example, calibrating with R filter photometry alone overestimates $\mbox{F}_{\lambda,cont}(\lambda_{H\alpha})$ of the $1$\,Myr old population at $z$=0 by 13\%. This systematic calibration error depends on the age of the stellar population as well as on its redshift, and propagates directly through \eqref{F_line_solved} to a systematic measurement error in $\mbox{F}_{line}$.

\section{The effect of errors in WNCR}
\label{sec:Err_Propagation}

In this section we derive the effect of the WNCR estimation error ($\Delta$WNCR) on the measurement errors of the equivalent width and flux of the line ($\Delta$EW and $\Delta \mbox{F}_{line}$).

Using \eqref{cps_line_N_solved} we derive $\left(\Delta \mbox{cps}_{N,line}\right)_{\tiny \mbox{WNCR}}$, the measurement error of $\mbox{cps}_{N,line}$ propagated from $\Delta$WNCR (i.e., $\Delta \mbox{cps}_{N,line}$ if all other sources of error are negligible). For reasons discussed above, the right-hand side factor of \eqref{cps_line_N_solved} typically introduces a small correction to the left-hand side factor. For error propagation purposes this correction can be ignored. Hence:

\begin{equation}
\begin{IEEEeqnarraybox*}{lCl}
\left(\Delta \mbox{cps}_{N,line}\right)_{\tiny \mbox{WNCR}} 
  & = &      \Delta \mbox{WNCR} \cdot \left| \frac{\partial \mbox{cps}_{N,line}}{\partial \mbox{WNCR}} \right| \\ 
  & \cong &  \Delta \mbox{WNCR} \cdot \frac { \mbox{cps}_{W} }{ \mbox{WNCR}^2 }
\end{IEEEeqnarraybox*}
\end{equation}

For the purpose of error analysis, we assume that the line's contribution to $\mbox{cps}_{W}$ is minor (i.e. $\mbox{cps}_{W} \cong \mbox{cps}_{W,cont}$). This is true in typical conditions, where the W-band width is significantly larger than the EW. 
Using \eqref{WNCR_def} with this assumption we obtain:

\begin{equation}
\left(\Delta \mbox{cps}_{N,line}\right)_{\tiny \mbox{WNCR}} 
  \cong \frac {\Delta \mbox{WNCR}}{\mbox{WNCR}} \cdot \mbox{cps}_{N,cont}
\label{Delta_CPS_N,line_WNCR_solved}
\end{equation}

From this and \eqref{F_line}, $\left( \Delta \mbox{F}_{line} \right)_{\tiny \mbox{WNCR}}$, the measurement error in $\mbox{F}_{line}$ propagated from $\Delta$WNCR, can be simply derived as: 

\begin{equation}
\begin{IEEEeqnarraybox*}{lCl}
\left( \Delta \mbox{F}_{line} \right)_{\tiny \mbox{WNCR}} 
   = 
\left(\Delta \mbox{cps}_{N,line}\right)_{\tiny \mbox{WNCR}} \cdot 
\frac{ \partial \mbox{F}_{line} } {\partial \mbox{cps}_{N,line}} \\
    \cong  
\frac {\Delta \mbox{WNCR}}{\mbox{WNCR}} \cdot 
\frac { \mbox{cps}_{N,cont} } 
      { \mbox{T}_{atm,N}(\lambda_{line}) \: \mbox{T}_{N}(\lambda_{line}) \: \mbox{R}_{\lambda}(\lambda_{line}) }
\end{IEEEeqnarraybox*}
\end{equation}

Using \eqref{cpsN_cont_est} and the assumption made in \eqref{EW_solved} yields:

\begin{equation}
\left( \Delta \mbox{F}_{line} \right)_{\tiny \mbox{WNCR}} 
  \cong 
\frac {\Delta \mbox{WNCR}}{\mbox{WNCR}}  \cdot
\frac { \int_0^\infty \mbox{T}_{N}(\lambda) \: d\lambda } { \mbox{T}_{N}(\lambda_{line}) } \cdot
\mbox{F}_{\lambda,cont}(\lambda_{line})
\label{F_line_err_solved}
\end{equation} 

$\left( \Delta \mbox{F}_{line} \right)_{\tiny \mbox{WNCR}}$ is therefore affected by the relative error in WNCR, the N-band effective width, and the underlying continuum flux. Note that $\left( \Delta \mbox{F}_{line} \right)_{\tiny \mbox{WNCR}}$ does not depend on the line's flux itself ($\mbox{F}_{line}$).
\\

Similarly, $\left( \Delta \mbox{EW} \right)_{\tiny \mbox{WNCR}}$, the measurement error in EW propagated from $\Delta$WNCR, can be derived from \eqref{cps_cont_line}, \eqref{EW_solved} and \eqref{Delta_CPS_N,line_WNCR_solved}:

\begin{equation}
\left( \Delta \mbox{EW} \right)_{\tiny \mbox{WNCR}} 
   \cong  
\frac {\Delta \mbox{WNCR}}{\mbox{WNCR}} \cdot
\frac {\int_0^\infty \mbox{T}_{N}(\lambda) \: d\lambda}{ \mbox{T}_{N}(\lambda_{line}) } \cdot 
\left( 1 + \frac { \mbox{cps}_{N,line} } {\mbox{cps}_{N,cont}} \right)
\label{EW_err_solved}
\end{equation}

The relative error, obtained by dividing this equation by \eqref{EW_solved}, is:

\begin{equation}
  \frac {\left( \Delta \mbox{EW} \right)_{\tiny \mbox{WNCR}}}
        {\mbox{EW}} 
\cong 
  \frac {\Delta \mbox{WNCR}}{\mbox{WNCR}} \cdot
  \left( 1 + \frac{\mbox{cps}_{N,cont}}{ \mbox{cps}_{N,line} } \right)
\label{rel_EW_err_solved}
\end{equation}

Since the last term in \eqref{EW_err_solved} is at least unity, it is evident that the relative error of WNCR multiplied by the N-band effective width is a lower limit of the uncertainty in EW:
\begin{equation}
  \Delta \mbox{EW} 
\geq 
  \frac {\Delta \mbox{WNCR}}{\mbox{WNCR}} \cdot
  \frac {\int_0^\infty \mbox{T}_{N}(\lambda) \: d\lambda}{ \mbox{T}_{N}(\lambda_{line}) } \cdot 
\label{EW_err_limit}
\end{equation}

Similarly, it is evident from \eqref{rel_EW_err_solved} that the relative error of WNCR sets a lower limit to the relative error of EW:

\begin{equation}
  \frac {\Delta \mbox{EW}}
        {\mbox{EW}} 
\geq 
  \frac {\Delta \mbox{WNCR}}{\mbox{WNCR}} \cdot
\label{rel_EW_err_limit}
\end{equation}

\section{WNCR Estimation Methods}
\label{sec:WNCR_MeasMethod}

For well-resolved galaxies, the continuum light can be directly measured from the galaxy itself. This can be done using the pixel-to-pixel method (B\"{o}ker et al.\ 1999; Rossa \& Dettmar 2000; Kennicutt et al.\ 2008; Finkelman et al.\ 2010) where $\mbox{cps}_{N}$ is plotted for each pixel as function of $\mbox{cps}_{W}$. To reduce the data scatter one may first reject a user-specified percentage of the brightest pixels, suspected to be foreground stars. 
Galaxies with smoothly varying light distribution, such as elliptical galaxies, can also be fitted with isophotal maps in both the W-band and N-band images. From these, one can plot the mean $\mbox{cps}_{N}$ vs. mean $\mbox{cps}_{W}$ for pairs of isophotes of the same semidiameter (e.g., Macchetto et al.\ 1996).

The resulting relation is expected to show most data points (presumably lacking line emission) aligning along a narrow straight line down to the noise level. The slope of this line yields the WNCR estimate. A smaller number of data points, which trace the H$\alpha$ emission, are expected to be above the line, whereas a small amount of scattered data points are also expected below the line. The latter trace either regions of excess continuum emission, or excess line absorption. 
Note that this WNCR estimate is affected by widespread diffuse emission and by underlying stellar H$\alpha$ absorption across the galaxy.

A hidden and problematic assumption in applying this method is that of a constant colour. The fitted regions may show colour variations caused by changes in stellar population across the galaxy and in the line-emission regions. These may be related to recent or ongoing SF episodes. Note that the derived WNCR estimate, which is affected by colour, does not represent the emission line region but is rather an average value across the galaxy. 
\\

An alternative method to estimate WNCR is by calculating the average $\mbox{cps}_{W}$ to $\mbox{cps}_{N}$ ratio of foreground stars (e.g., Pogge, Owen \& Atwood 1992; Rand 1996; Koopman, Kenney \& Young 2001; James et al.\ 2004). This method allows measuring line emission even if the continuum-dominated regions of the target cannot be properly sampled.

The main assumption is that the foreground stars represent the mix of stars in the target galaxy, thus matching on average the spectral distribution of its continuum. 
In order to represent a continuous distribution function of stellar types, as many foreground stars as possible should be measured. However, regardless of how numerous, foreground stars represent the local population of the region in the Milky Way through which the observation is made, rather than the stellar population in the target galaxy. These two populations may be significantly different.

This method is more useful for target galaxies that are sufficiently distant to allow the N-band to transmit their rest-frame H$\alpha$ while blocking the local H$\alpha$ (6563$\mbox{\AA}$) of Milky Way foreground stars. 
If this is not the case, applying this method implies also the assumption that the fraction of foreground stars with strong Balmer line absorption or emission (typically O, B and A stars) is insignificant. 
This assumption typically holds for observations in mid to high Galactic latitudes. At these latitudes Galactic foreground stars between the observer and a target tend to be of types later than A due to the age of the halo and the shape of the stellar initial mass function.
\\

We note that an alternative set of bands may be used, in which an off-line narrow band replaces the wide W-band (e.g. Almoznino \& Brosch 1998; van Zee 2000; Dale et al.\ 2010). If the central wavelengths of the off-line and on-line bands are close enough, the WNCR uncertainty due to colour effects may be reduced. 
However, this comes at the expense of increased random errors due to the lower depth of images obtained with limited telescope time.
An alternative approach utilises two medium to wide off-line bands flanking the H$\alpha$ line (e.g. MacKenty et al. 2000; Lin et al. 2003; Calzetti et al. 2004; Ly et al. 2011). This approach reduces colour effects, while having a smaller impact on image depth.

\section{Colour Effects }
\label{sec:ColourEffects}

Colours of the studied emission-line regions may differ from those of the foreground stars or those of the continuum-dominated galactic regions used for estimating WNCR. As a result, the estimated WNCR may deviate from its value in the studied region. Furthermore, since colours typically vary within a galaxy, different parts of a target galaxy may have different WNCR values. In an attempt to improve accuracy, the value of WNCR is sometimes adjusted subjectively to obtain a desirable continuum subtracted image (Rand 1996; James et al.\ 2004; Kennicutt et al.\ 2008).

Although the uncertainties in estimating WNCR are typically of few percent, these can result in a much higher uncertainty in the measured H$\alpha$ flux, especially for low EW values (Macchetto et al.\ 1996; Hameed \& Devereux 1999; Fig. 7 in Koopmann et al.\ 2001). For example, for a $5\%$ error in WNCR and an N-band effective width of 100$\mbox{\AA}$, the relative error of EW would be at least 5\% (eq. \ref{rel_EW_err_limit}), while the absolute error of EW would be at least 5$\mbox{\AA}$ (eq. \ref{EW_err_limit}). For low SF rates the latter can be comparable to the measured EW.

In order to estimate the variation of WNCR with colour we model observations of different objects with recession velocities up to 30000\,km s$^{-1}$ (z=0.1). For each object we simulate observations through a Johnson-Cousins R filter and a narrow-band filter, 50$\mbox{\AA}$ wide and centred on the redshifted H$\alpha$ line.
To simulate foreground stars we use various spectra of Main Sequence (MS) and off-MS stars, taken from the Bruzual-Persson-Gunn-Stryker atlas (available at ftp://ftp.stsci.edu/cdbs/grid/bpgs/). The continua of different parts of target galaxies are represented by redshifted spectra of single stellar population (SSP) models at different ages in the range $10^6$-$10^{10}$\,yr. These star-bursts are characterised by a Salpeter initial mass function (IMF) with a lower and upper mass of 0.1 and 125\,M$_\odot$, respectively. The data are taken from the Bruzual-Charlot atlas (available in ftp://ftp.stsci.edu/cdbs/grid/bc95/templates/).

\begin{figure*}
\begin{tabular}{cc}
{\includegraphics[width=6cm,trim=0mm 0mm 0mm 0, clip]{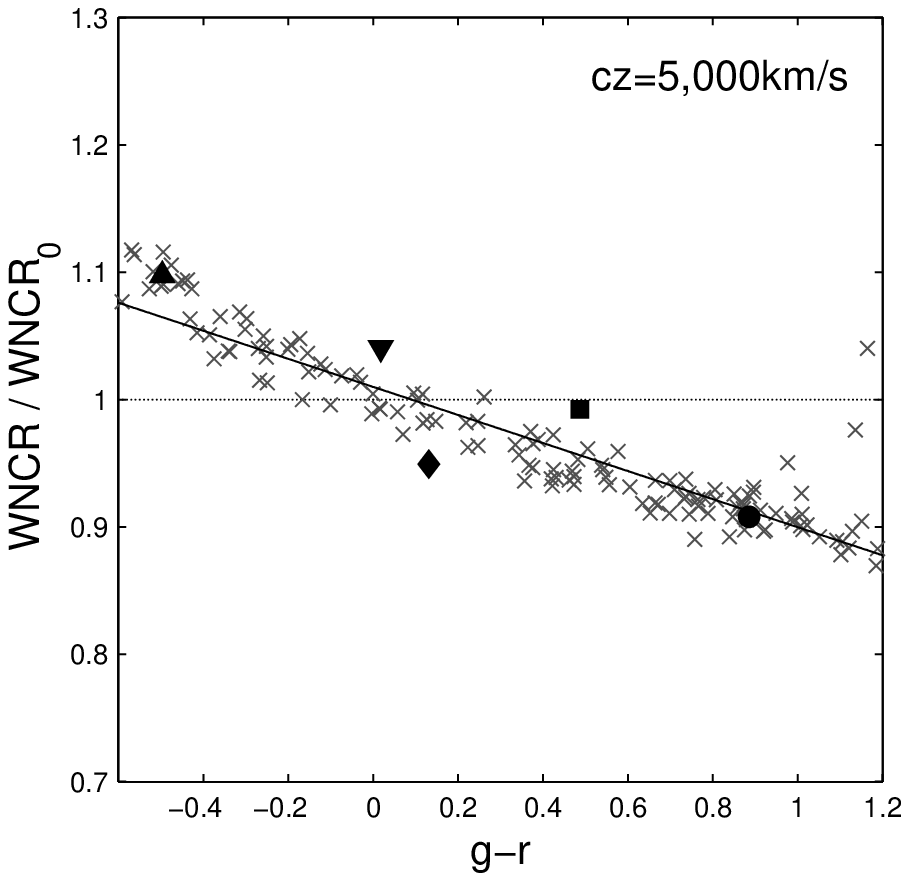}} & {\includegraphics[width=6cm,trim=0mm 0mm 0mm 0, clip]{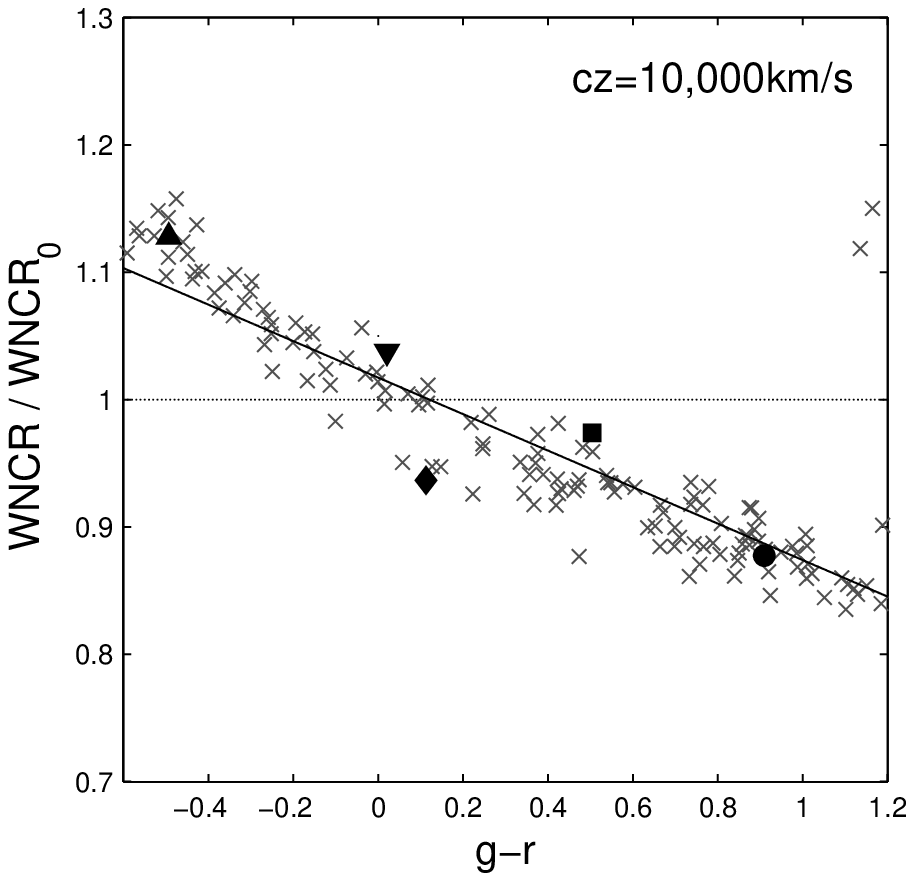}}\\
{\includegraphics[width=6cm,trim=0mm 0mm 0mm 0, clip]{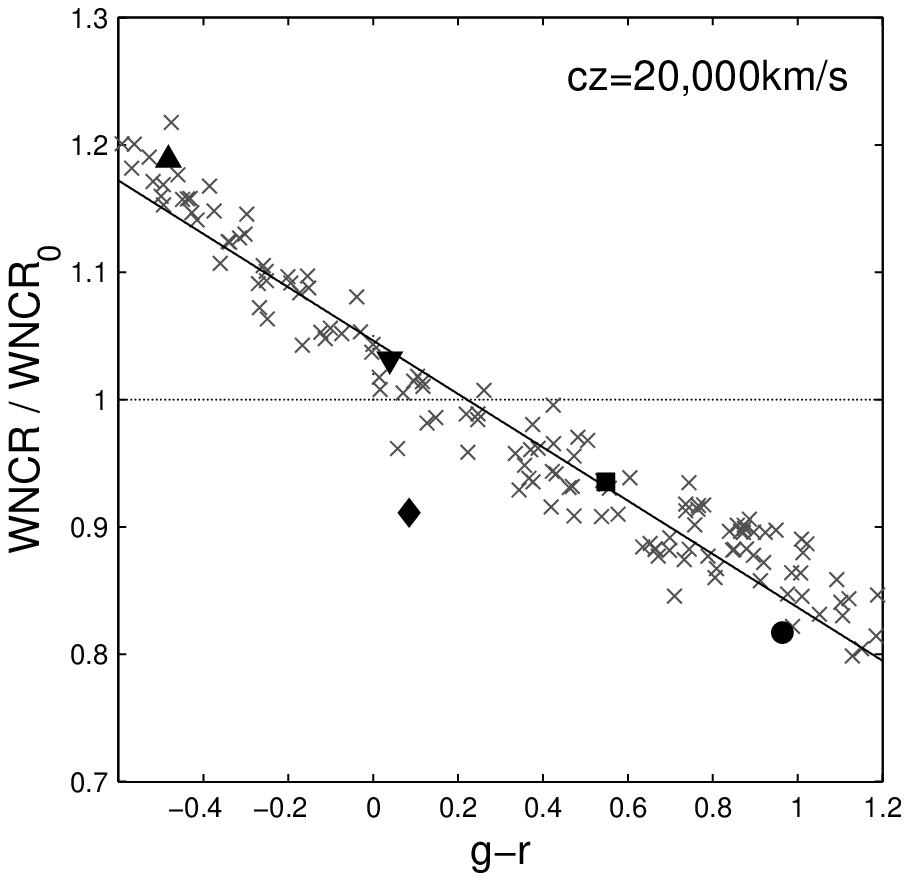}} & {\includegraphics[width=6cm,trim=0mm 0mm 0mm 0, clip]{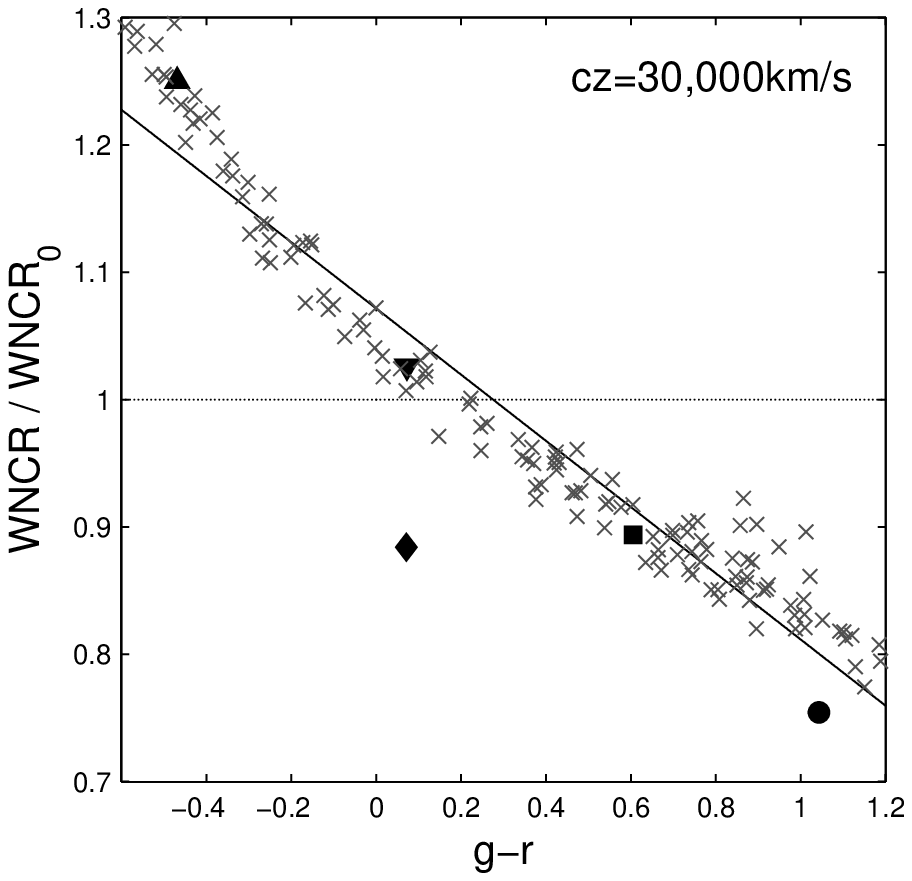}}
\end{tabular}
\caption
{The theoretical WNCRs measured for a set of objects with a set of narrow-band filters. 
Foreground stars of various spectral types and luminosity classes are represented by $\times$, while redshifted SSPs evolving with time are marked by filled shapes ($\CIRCLE$ - $10^{10}$\,yr, $\blacksquare$ - $10^9$\,yr, $\blacktriangledown$ - $10^8$\,yr, $\blacklozenge$ - $10^7$\,yr, $\blacktriangle$ - $10^6$\,yr).
The set of filters sample the rest-frame H$\alpha$ line in galaxies with recession velocities of 5000, 10000, 20000 and 30000\,km s$^{-1}$. The WNCR values measured with each narrow-band filter are normalized by the corresponding WNCR$_0$ value calculated for a constant $f_\nu$ spectrum (zero colour in the AB magnitude system). The solid lines show linear fits of the foreground stars. 
\label{f:WNCRHa}
}
\end{figure*}

Using the spectra of these objects, the filters' transmittance curves, and the responsivity of a typical electro-optical system, we calculate $\mbox{cps}_{W,cont}$ and $\mbox{cps}_{N,cont}$ and from it the theoretical WNCR for each object (equations \ref{cps_cont_line_detailed} and \ref{WNCR_def}).
The calculated WNCRs for each narrow-band filter are plotted vs.\ the computed SDSS $g$-$r$ colour of each object (Figure \ref{f:WNCRHa}). For each narrow-band filter WNCR values are normalized using the WNCR of a theoretical object with constant $f_\nu$ (zero colour in the AB magnitude system). Linear fits for foreground star points are also shown in each plot. 

As can be seen in Figure \ref{f:WNCRHa}, WNCR of the foreground stars depends almost linearly on $g$-$r$. WNCR of the simulated target galaxy parts seem to follow the curve of the foreground stars with the exception of two points ($10^7$\,yr old SSP population with recession velocities of 20000 and 30000\,km s$^{-1}$).

Based on this, we suggest the following method to obtain a more accurate estimate of WNCR. A colour measurement of the objects in the imaged field should be obtained either by imaging in a second wide band, or by using public databases. WNCR values should be measured for each reference object (foreground star or part of the target galaxy with no line flux) by dividing the counts in W by the counts in N (equation \ref{WNCR_def}). A linear relation should then be fitted to the measured (colour,WNCR) points. Finally, this linear relation should be used to obtain a WNCR estimate for each part of interest in the target galaxy as function of its colour.

\begin{figure*}
\begin{tabular}{cc}
\includegraphics[width=6cm,trim=0mm 0mm 0mm 0, clip]{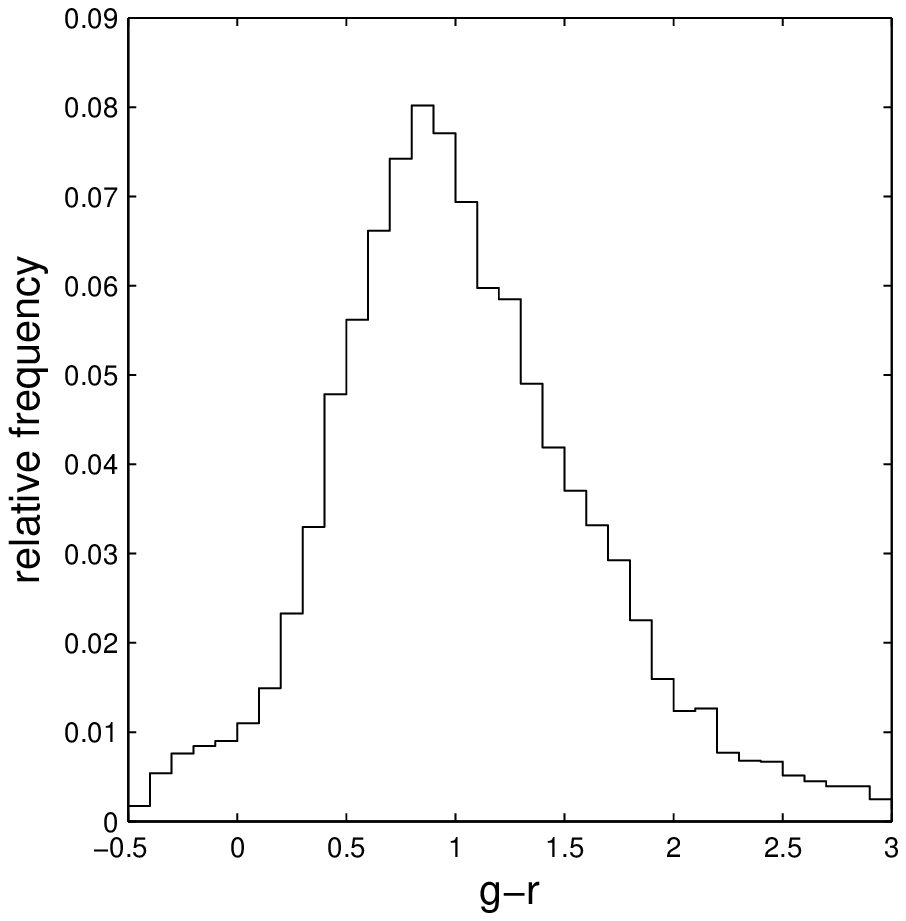} &
\includegraphics[width=6cm,trim=0mm 0mm 0mm 0, clip]{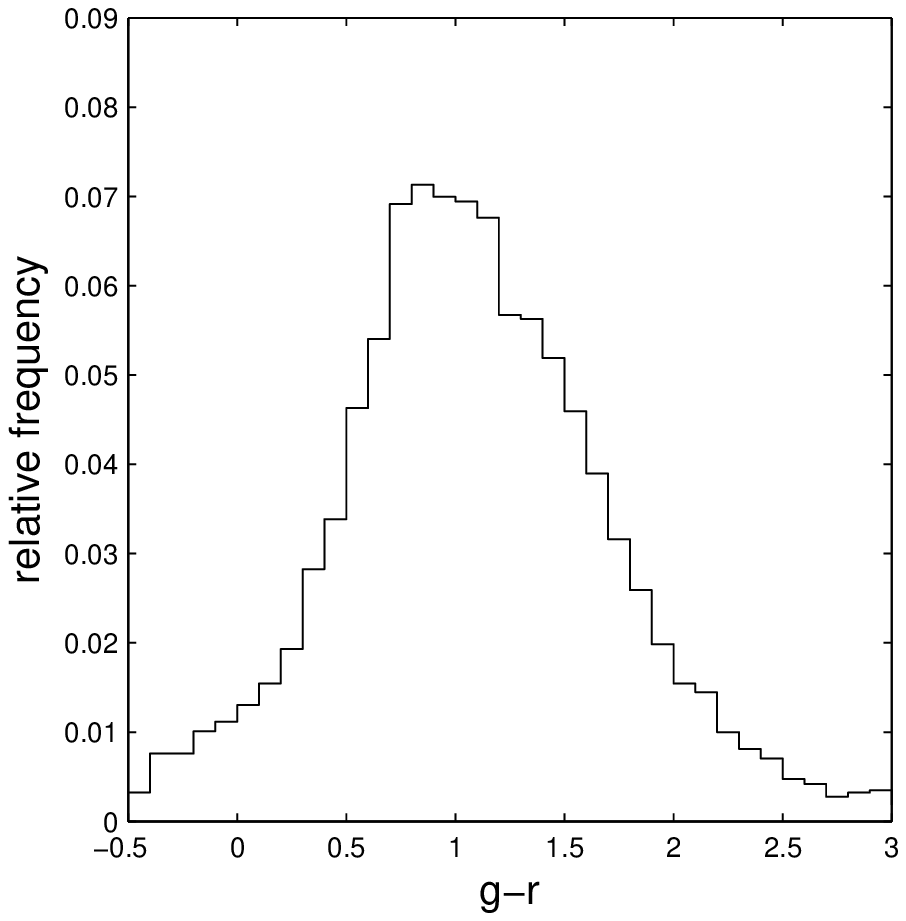}
\end{tabular}
\caption
{ The distribution of the $g$-$r$ colour of stars in two regions in the sky. The regions were arbitrarily selected to include all stars in a circular aperture of 30\,arcmin around two arbitrary nearby galaxies (NGC 5363 and NGC 7625). 
 \label{f:stars}
}
\end{figure*}

Let us now estimate the possible errors in WNCR, when obtained without colour information. First, we estimate the normalized WNCR value obtained using the foreground stars method. For this we measure the colour of foreground stars around arbitrary nearby galaxies. This is done by collecting data from the SDSS archive for regions around these galaxies, each with a radius of 30\,arcmin, and containing $\sim$10,000 objects classified as stars. The $g$-$r$ colour histograms for two such regions are presented in Figure \ref{f:stars}. 
Both histograms show a distribution that peaks at $g$-$r$ of $\sim$0.8, corresponding to lower main sequence and giant branch stars. Since the average colour is also of a similar value, we adopt 0.8 as a typical $g$-$r$ colour for foreground stars around galaxies.

The error in WNCR can be estimated for the different recession velocities of Figure \ref{f:WNCRHa} by comparing the WNCR value of foreground stars at $g$-$r$\,=\,0.8 with the WNCR values of the simulated target galaxy parts (the redshifted stellar populations). When, for example, measuring a starburst galaxy dominated by a young stellar population (10$^6$\,yr) with foreground stars of $g$-$r$\,=\,0.8, WNCR will be underestimated by up to $\sim$30\% depending on the galaxy's redshift (for cz $\leq 30000$\,km s$^{-1}$).
This means underestimating the H$\alpha$ EW by a similar fraction (equation \ref{rel_EW_err_limit}), and underestimating $\mbox{F}_{line}$ by $\sim$0.3 of the continuum flux within the N-band (equation \ref{F_line_err_solved}). 
For older underlying stellar populations the error in WNCR would be smaller. For 10$^9$\,yr the underestimation will reach 8\% at 10000\,km s$^{-1}$, while for 10$^{10}$\,yr WNCR will be overestimated by up to $\sim$10\%.

Estimating WNCR using galaxy parts believed not to form stars may also be problematic, since it does not take into account the possible colour variations across the galaxy and the expected colour change due to star formation in the H$\alpha$ emitting regions. Let us consider a case where a 10$^6$\,yr old stellar population is observed within an elliptical galaxy dominated by a 10$^{10}$\,yr old population. The flux from the young population is expected to overwhelm the flux from the old population, even if it is only a small fraction of the entire star population. Therefore, the measured WNCR will correspond to redder objects, i.e. would be underestimated by up to $\sim$40\% depending on redshift (for cz $\leq 30000$\,km s$^{-1}$).

These errors in WNCR propagate through $\mbox{F}_{line}$ to SF measurements. Since the errors depend on colour properties of the galaxies, a bias in the colour properties of a sample may result in a bias in the SF properties concluded by a survey. 

This applies also to surveys that measure cosmic SF rate at low redshifts (low z points in the Madau plot). Since star forming galaxies tend to be bluer than foreground stars, results may be biased. As discussed above, such a survey that uses foreground stars and no colour information may result in the underestimation of WNCR by tens of percents. Such underestimation, which increases with redshift, will result in the underestimation of the cosmic SF rate.

\section{Conclusions}
We identified an important error in estimating extragalactic H$\alpha$ emission, originating from commonly used procedures of continuum subtraction.
We showed that when measuring an emission line by observing through a narrow-band (N) and a wide-band filter (W), where the N band contains the emission-line, and the W band may or may not contain it:

\begin{enumerate} 
 \item The Equivalent Width, EW, can be obtained without the need for photometric calibrations using \eqref{cps_line_N_solved} and \eqref{EW_solved}. To improve accuracy:
 \begin{itemize}
  \item the N filter should have a very narrow band and negligible ``out of band leaks'';
  \item the exposures of the W and N bands must be taken in very similar atmospheric conditions and air-masses, or with measured atmospheric transmittance at the line's wavelength [$\mbox{T}_{atm,W}(\lambda_{line})$ and $\mbox{T}_{atm,N}(\lambda_{line})$]. 
 \end{itemize}

 \item An accurate measurement of the line flux, $\mbox{F}_{line}$, can be obtained by photometric calibrations in the N-band, if it is sufficiently narrow to assume that the responsivity, $\mbox{R}_{\lambda}$, and the atmospheric transmittance, $\mbox{T}_{atm,N}$, are constant within it. Alternatively, if calibrated measurements of the magnitude in several wide-bands are obtained, a fit can be used to estimate the continuum flux under the line, $\mbox{F}_{\lambda_{line},cont}$, from which $\mbox{F}_{line}$ can be obtained. Using calibrations in the W-band alone may lead to systematic errors of over 10\%. \\

 \item An accurate estimation of WNCR, the W to N band ratio of count rates contributed by the continuum (equation \ref{WNCR_def}), is crucial for accurate emission line measurements:
 \begin{itemize}
  \item The uncertainty in EW is equal to or greater than the relative error in WNCR multiplied by the N-band effective width (equation \ref{EW_err_limit});
  \item The relative uncertainty in EW is equal to or greater than the relative error in WNCR (equation \ref{rel_EW_err_limit});
  \item The uncertainty in the emission-line flux, $\mbox{F}_{line}$, is equal to or greater than the relative error in WNCR multiplied by the underlying continuum flux and by the N-band effective width (equation \ref{F_line_err_solved}). 
 \end{itemize}

\item The value of WNCR may vary significantly with the colour of the measured object. For typical H$\alpha$ measurements at redshifts cz $\leq 30000$\,km s$^{-1}$ the dependence on $g$-$r$ was found to be almost linear. Where foreground Milky Way stars are used for estimating WNCR, colour effects may lead to an underestimation as great as 30\%, or in other cases overestimation by as much as 10\%. Where regions in the target galaxy itself are used for the estimation, WNCR may be underestimated by as much as 40\%. \\

\item We suggest estimating WNCR as a linear function of target colour, rather than as a constant value. This will reduce errors, and will make the measurement procedure more objective. \\

\item The errors described above can bias the results of a study of a class of galaxies by systematically underestimating or overestimating the emission-line flux. Cosmic SF rate measurements at low redshifts may also be biased by this effect. If not corrected properly, such measurements may underestimate the low z points in the Madau plot. 

\end{enumerate}

\section*{Acknowledgements}

Funding for the SDSS and SDSS-II has been provided by the Alfred P. Sloan Foundation, the Participating Institutions, the National Science Foundation, the U.S. Department of Energy, the National Aeronautics and Space Administration, the Japanese Monbukagakusho, the Max Planck Society, and the Higher Education Funding Council for England. The SDSS Web Site is http://www.sdss.org/.

\label{lastpage}

\end{document}